\documentclass[fleqn,10pt]{wlscirep}
\usepackage[utf8]{inputenc}
\usepackage[T1]{fontenc}
\usepackage{mathtools}
\usepackage{caption}
\usepackage{hyperref}
\usepackage{lineno}
\title{The Influence of Biomedical Research on Future Business Funding: Analyzing Scientific Impact and Content in Industrial Investments}

\author[1,*]{Reza Khanmohammadi}
\author[2]{Simerjot Kaur}
\author[2]{Charese H. Smiley}
\author[3]{Tuka Alhanai}
\author[2]{Ivan Brugere}
\author[2]{Armineh Nourbakhsh}
\author[1]{Mohammad M. Ghassemi}
\affil[1]{Michigan State University, Computer Science and Engineering, East Lansing, USA}
\affil[2]{JPMorgan Chase, Artificial Intelligence Research, New York, USA}
\affil[3]{New York University Abu Dhabi, Computer Engineering, Abu Dhabi, UAE}
    
\affil[*]{khanreza@msu.edu}

\begin{abstract}
This paper investigates the relationship between scientific innovation in biomedical sciences and its impact on industrial activities, focusing on how the historical impact and content of scientific papers influenced future funding and innovation grant application content for small businesses. The research incorporates bibliometric analyses along with SBIR (Small Business Innovation Research) data to yield a holistic view of the science-industry interface. By evaluating the influence of scientific innovation on industry across 10,873 biomedical topics and taking into account their taxonomic relationships, we present an in-depth exploration of science-industry interactions where we quantify the temporal effects and impact latency of scientific advancements on industrial activities, spanning from 2010 to 2021. Our findings indicate that scientific progress substantially influenced industrial innovation funding and the direction of industrial innovation activities. Approximately 76\% and 73\% of topics showed a correlation and Granger-causality between scientific interest in papers and future funding allocations to relevant small businesses. Moreover, around 74\% of topics demonstrated an association between the semantic content of scientific abstracts and future grant applications. Overall, the work contributes to a more nuanced and comprehensive understanding of the science-industry interface, opening avenues for more strategic resource allocation and policy developments aimed at fostering innovation.
\end{abstract}
\begin{document}
% \linenumbers

\flushbottom
\maketitle
% * <john.hammersley@gmail.com> 2015-02-09T12:07:31.197Z:
%
%  Click the title above to edit the author information and abstract
%
\thispagestyle{empty}
\begin{figure}[t]
  \centering
  \includegraphics[width=1\linewidth]{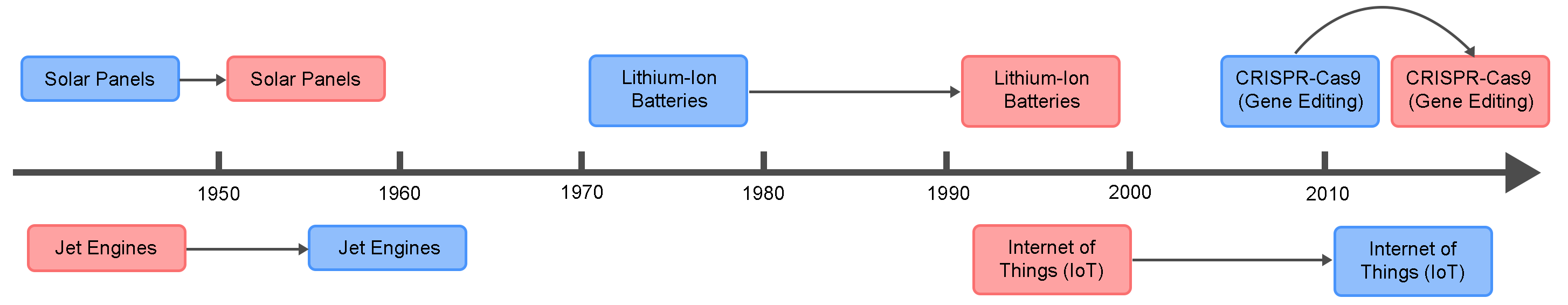}
  \caption{The bidirectional flow of breakthroughs between academia and the industry. Above the timeline, blue boxes represent scientific discoveries that have later catalyzed commercial applications, as indicated by the arrows leading to red boxes. Below the timeline, red boxes mark industry innovations that have subsequently inspired scientific exploration and advancement, traced back to blue boxes. The interchanging paths underscore the dynamic exchange between scientific inquiry and industrial application across different fields.
  \label{fig:cover-figure}}
\end{figure}
\section*{Introduction}
It is widely believed that scientific innovation influences present-day or future industrial activities; this influence is especially pronounced in the biomedical and health sciences, where scientific validation is often a pre-requisite to commercial translation \cite{Gopukumar}. CRISPR-Cas9 \cite{Jinek} is one recent example, illustrated in Figure \ref{fig:cover-figure}, that demonstrates how scientific innovation can influence the direction of industrial activity. Originating from basic biological research into the immune systems of bacteria, CRISPR-Cas9 has impacted the biomedical industry by introducing a more efficient and precise method for gene editing; this in-turn has led to numerous downstream industrial applications, including more effective gene therapies \cite{Mounadi}. However, the magnitude and time-scale of a given scientific innovation's impact on downstream industrial activities can vary dramatically by research topic, and a host of other complex forces including government regulation, economic conditions, and societal perceptions. For instance, research on the role of Telomere decomposition in the aging process has provided important insights into how and why biological organisms age \cite{shammas2011telomeres}, but has not (yet) resulted in downstream industrial applications. To date, a detailed characterization of the magnitude and time-scale of scientific activities' impact on downstream industrial activities is missing --- addressing this gap is the primary objective of this research paper. 

Understanding the relationship between scientific innovation and industrial activity has substantial practical implications. Identifying scientific areas with a significant impact on the industry will enable businesses to anticipate future trends and adapt investment strategies accordingly. Conversely, when researchers are familiar with the commercial implications of their research, they are more likely to conduct research that has a high chance of being translated into practical applications. For example, knowing the industrial relevance of developing more effective antibiotics can guide biomedical researchers to focus their efforts on this area, potentially resulting in improved public health outcomes and opportunities for economic growth in the pharmaceutical sector.

\subsection*{Research Questions} \label{sec:rques}
Our key objective is to clarify how \textit{present} areas of scientific innovation was associated with \textit{future} small business commercialization activities and funding within the same innovation areas. More specifically, this study aims to answer two research questions:

\begin{enumerate}
\item For a given topic in the biomedical sciences, can the \textit{historical impact} of scientific papers be a leading indicator of \textit{future funding} allocations to small businesses active in those topics?

\item For a given topic in the biomedical sciences, can the \textit{historical content} of scientific abstracts be a leading indicator of the \textit{future content} of innovation grant applications for small businesses active in those topics?
\end{enumerate}

If these research questions can be answered in the affirmative, it implies several opportunities for additional research and development. For example, if historical trends in scientific activities are associated with future funding allocations to small businesses working on associated topics, then it may be possible to forecast future (unknown) industrial trends based on current (known) scientific activities. Insofar as this forecasting can be performed with fidelity, this understanding can guide resource allocation, investment strategies, and inform policy developments intended to foster innovation. 

\subsection*{Related Work}
The relationship between scientific innovation and industry is crucial for technological and economic progress. The U.S.'s proposed \$191 billion funding for research and development in 2023 \cite{N2023} shows the importance of science in driving industrial growth. To understand this complex relationship, researchers often use bibliometric and patent analyses, which provide objective views on the ongoing interaction between science and industry.

\subsubsection*{Bibliometric Analysis} \quad
Bibliometric analysis is a method often used to delve into scientific literature and detect evolving trends, providing a way to illustrate the connections between science and industry \cite{Jürgens}. This approach has been utilized in various studies to investigate aspects such as university-industry collaborations \cite{Skute}. It has also been used to examine the scientific output in the publishing industry, identifying leading academic publications, top authors, and primary research countries \cite{García}. Notably, these studies highlighted the US, UK, Spain, and China as leading nations in scientific output. Further research identified central themes like cyber-physical systems and cloud computing in the Industry 4.0 research field \cite{COBO2018364}. Thus, bibliometric analysis facilitates a deeper understanding of scientific trends, their potential applications, and how they shape and are shaped by industrial development and socio-economic influences. 

\subsubsection*{Patent Analysis} \quad
While bibliometric analysis provides insights into scientific trends and their influence on industry, it alone cannot fully capture the complexity of industrial activities their evolution. To complement this approach, patent analysis offers a focused examination of industry-related intellectual property and technological advancements \cite{KRESTEL2021102035}. For example, a study on Carbon Capture, Utilization, and Storage (CCUS) technology revealed its rapid development since 2013, with patents concentrated in China, the US, and Japan, particularly in the energy and electricity sectors \cite{su15043484}. Moreover, patent analysis enhances our understanding of industry-focused aspects. For instance, a scientometric study on Smart Cities showed that research predominantly focuses on social aspects, while related technologies emphasize specific technical solutions, often overlooking the role of citizens \cite{PULIGA2023}. Wang and Li\cite{Zexia} also highlighted the significant impact of high-quality academic research on patent development in the realm of nanotechnology, using data from nano patents and their associated citations. They also shed light on the variety in citation patterns, influenced by factors like organizational type and origin of knowledge, and suggest that a broader scientific scope does not necessarily translate into higher patent quality. By incorporating patent analysis alongside bibliometric analysis, researchers can gain a more comprehensive view of the intricate relationship between scientific progress and industrial evolution \cite{Chakraborty2020PatentCN}.

\subsubsection*{Empirical Insights from PubMed papers and SBIR awards}
PubMed, a renowned platform for biomedical research, has been at the forefront of capturing evolving scientific trends. Recent studies on this platform cover a wide array of specific biomedical topics such as Kawasaki disease \cite{Tan2022}, COVID-19 \cite{Khalid}, protein engineering \cite{Mardikoraem}, and spine surgery \cite{Maghrabi}, as well as broader public health topics such as water quality \cite{Chandrasekar} and vaccine clinical trials \cite{Mohana}. These in-depth analyses offer a window into the current state and gaps in medical research, highlighting areas of priority. The SBIR program, on the other hand, offers specific insights into technological advancements and commercialization strategies within the industry sector. Audretsch et al.\cite{Audretsch2019KnowledgeBK}'s exploration of SBIR-awarded projects between 1992 and 2001 underscores the value of intertwining university expertise with industry-driven goals. In addition to enhancing the quantity of scientific papers, this intersection enhances the richness of technological advances. Accordingly, Hayter and Link\cite{Hayter}'s analysis of 1,180 SBIR-endorsed firms emphasizes the complementary role of patenting and publishing as intertwined components of technology commercialization. Despite their valuable individual contributions, PubMed studies and SBIR projects offer largely untouched opportunities to explore the science-industry relationship more comprehensively.

\subsubsection*{Gaps in the Literature} \quad
Despite the insightful findings of these studies, several gaps persist in the literature:
\begin{itemize}
\item The diversity across the number of topics addressed in these studies has been rather limited.
\item While patents provide a glimpse into technological advancements, they do not fully capture the nuances of businesses that operate on these technologies, nor do they directly reflect financial implications.
\item The semantic content of scientific and industrial activities has not been adequately accounted for in previous research.
\item The temporal aspect of the evolution, which could be viewed as a time series analysis, remains largely unexplored, as does the time it takes for the impact of science to materialize in industry.
\end{itemize}

\subsection*{Contributions of this work}
Our contributions in this work provide a more comprehensive understanding of the science-industry interface, helping to bridge existing gaps in the literature. We provide:

\begin{itemize}
    \item \textbf{Broad assessment of biomedical research topics:} 
    We explore the effect of scientific innovation on industrial activities within 10,873 biomedical topics; this is the most comprehensive exploration of its kind, to the best of our knowledge.
    
    \item \textbf{Accounting for topical taxonomic relationships:} 
    The topics investigated are hierarchically structured; thus, we account for the taxonomic relationship and rank of the topics. By accounting for these taxonomic properties, our analysis also clarifies if scientific innovation influences very specific industrial activities (e.g. CRISPR-Cas9), or broader industrial trends (e.g. Genetic Phenomenon).

    \item \textbf{Assessment of temporal effects and impact latency:} 
    We examine the temporal evolution of scientific innovation's influence on industry, analyzing the latency of its impact from 2010 to 2021.

    \item \textbf{Diverse analytical approaches:} 
    We employ various methods to understand the science-industry connection, including correlation and causality studies, as well as analyzing content overlaps between scientific and industrial texts.
    
\end{itemize}
We hope that our study bridges the gap that exists in the literature between analyzing scientific research and probing the trends in industry investments.

\section*{Methods}
In the Methods section, we outline the data sources and analytical approaches employed to address the research questions outlined earlier. We detail the bibliometric and industrial innovation datasets used, describe the taxonomy for categorizing biomedical topics, and present our methodologies for assessing the relationships between scientific interests and industry funding, as well as the semantic content of academic and industrial texts.

% In this section we provide an overview of the data and analytic approach used to answer the research questions from Section "\hyperref[sec:rques]{Research Questions}". A succinct overview of this section follows: In Subsection "\hyperref[sec:data]{Data}", we describe the biobliometric data \cite{pubmed} and industrial innovation data \cite{sbir} used in this study. In Subsection "\hyperref[sec:meshlabeling]{Selected Topic Taxonomy}" we describe the taxonomy that was used to assign papers and industrial innovation activities to biomedical topics. In Subsection "\hyperref[sec:freqimpactanalysis]{Methods for Research Question 1}", we describe the method used to explore the association between scientific interest in a given topic, and the annual funding granted to small businesses performing activities to commercialize products related to those topics. In Subsection "\hyperref[sec:contextanalysis]{Methods for Research Question 2}", we describe the method used to explore the association between the semantic contents of papers in a given topic, and the semantic contents of grant applications of small businesses working on those topics.

\subsection*{Data} \label{sec:data}

\subsubsection*{Data Sources}
All data for this study were publicly available; they were sourced from: (1) the abstracts and metadata of 10,928,078 scientific publications in PubMed \cite{pubmed} and, (2) the project abstracts and award amounts of 63,488 Small Business Innovation Research (SBIR) grants \cite{sbir}. Both assets were relevant to the time period spanning 2010 to 2021. %For future discussion of these data, we define \(P_{2010:2021}\) and \(S_{2010:2021}\) as PubMed and SBIR abstracts from 2010 to 2021, respectively. 

%The PubMed dataset consists of abstracts from the biomedical literature and has an average length of 205 words, while the SBIR dataset represents research and development projects in the industry and has an average abstract length of 251 words. 

\subsubsection*{Selected Topic Taxonomy} \quad \label{sec:meshlabeling}
To analyze the impact of scientific activity on industrial innovation across a consistent set of topics, we utilized the Medical Subject Headings (MeSH) taxonomy (\urlstyle{same}\url{https://www.ncbi.nlm.nih.gov/mesh/}). MeSH is a controlled and hierarchically organized vocabulary produced by the National Library of Medicine that facilitates indexing, cataloging, and searching for biomedical and health-related information. MeSH terms are organized in a tree-like structure, with more general terms at higher levels and more specific terms at lower levels.

The MeSH ontology consists of more than 29k terms, a large proportion of which did not occur in any SBIR abstract from 2010 to 2021. This absence can be largely attributed to the high specificity of many MeSH terms, which do not align with the more general, high-level language and vocabulary typically employed in SBIR abstracts. For instance, "Vibrio vulnificus" is a MeSH term that represents a specific species of bacteria. Given its high degree of specificity, it is unlikely to appear in SBIR abstracts. However, its higher-level parent term in the MeSH hierarchy, "Bacteria", is much more general, and therefore more likely to be appear in SBIR abstracts. To create a more manageable and relevant dataset for our investigation, we reduced the number of MeSH terms to 10,873. This revised subset only included topics that were present at least once in the SBIR abstracts between 2010 and 2021, providing a more effective scale for exploring the science-industry relationship. 

\subsubsection*{Topic Annotation Approach}
\begin{figure}[t]
  \centering
  \includegraphics[width=0.85\linewidth]{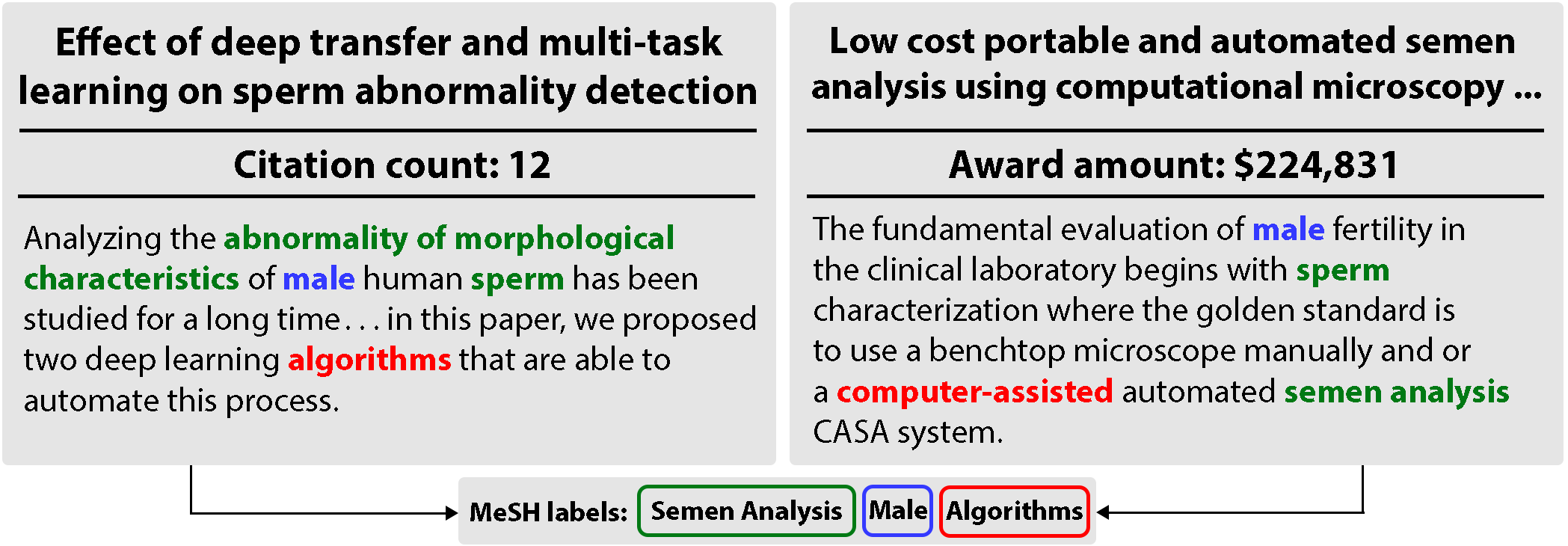}
  \caption{Illustration of shared MeSH term labels between a PubMed paper \protect\cite{ABBASI2021104121} and an SBIR project description (\urlstyle{same}\url{https://www.sbir.gov/node/1327383}). 
  \label{fig:mesh_labeling}}
\end{figure}

The papers collected from PubMed include human-generated MeSH annotations; however, the SBIR awards were not annotated for their topical contents (MeSH or otherwise). We generated the missing MeSH annotations for the SBIR awards by applying ScispaCy \cite{Neumann_2019} to the SBIR abstracts, extracting Unified Medical Language System (UMLS) \cite{Bodenreider} labels with a confidence score exceeding 90\%, and converting the UMLS topics to their corresponding entires within MeSH using the UMLS  API (\urlstyle{same}\url{https://documentation.uts.nlm.nih.gov/rest/home.html}). An example scientific paper and industrial project are shown in Figure \ref{fig:mesh_labeling}, along with their respective shared MeSH terms.

\subsection*{Methods for Research Question 1} \label{sec:freqimpactanalysis}
For a given topic in the biomedical sciences, the goal of our first research question is to understand if the the historical incidence or impact of scientific papers (measured by citations) can be a leading indicator future funding allocations to small businesses working on the same topics. To explore the relationship between science and industry for our first research question, we develop metrics that capture the incidence and impact of scientific publications as well as the investment trends in SBIR grants. We apply these metrics to both scientific publications and industrial data within each MeSH topic and track them over time. This allows us to examine the association between scientific advancements and subsequent funding allocations to small businesses within the same topics.

For each topic, we represent the scientific and industrial data in two ways: (1) as the normalized annual frequencies of published PubMed papers and awarded SBIR grants related to that topic and, (2) as the normalized citation counts of published PubMed papers and cumulative funding awarded through SBIR grants related to that topic. Formal details of the  approach are detailed below.\\

\subsubsection*{Representing Data as Paper and SBIR Award Frequency} \quad \label{sec:freq}
We represent scientific activity ( $f^P$ ) as a min-max normalized timeseries of the \textit{total annual number of papers} published on a given topic ( $m$ ) or its children topics ( $m_c$ ) in MeSH. We represent industrial activity ( $f^S$ ) as a min-max noramalized timeseries of the \textit{total annual number of SBIR grants} awarded to small businesses working on those topics, or their children topics in MeSH. In Equation 1, we formally denote how $f^P$ and $f^S$ signals are generated from the set of all PubMed abstracts $P(t)$ and SBIR abstracts $S(t)$:
\begin{equation} \tag{1}
f^X(t, m) = \frac{1}{{N^x(t)}} \sum_{j \in X(t)} H^j(m) 
\end{equation}
\noindent Where: $X \in \{P, S\}$; $t \in \mathbb{Z}$, $2010 \leq t \leq 2021$; $N^x(t)$ represents the  total number of (PubMed or SBIR) abstracts in year $t$ and
\begin{equation} \tag{2}
H^j(m) = \mathbf{Heaviside} \left( -1 + \sum_{i \in \{m,m_c\}}  \delta(j, i) \right) 
\end{equation}
\noindent Where: $\delta(j,i)$ is a function that returns 1 if abstract $j$ is on topic $i$ and 0 otherwise. $f^{P}$ and $f^{S}$ were further normalized within each topic (i.e. across time) using Min-Max scaling (i.e. rescaling the values of the signals into the range of 0 to 1). We denote $\tilde{f}^{P}$ and $\tilde{f}^{S}$ as the Min-Max normalized versions of $f^{p}$ and $f^{s}$s respectively: 
 \begin{equation} \tag{3}
\tilde{f}^{X}(t,m) = \frac{f^{X}(t,m) - \min f^{X}(:,m)}{\max f^{X}(:,m) - \min f^{X}(:,m)}
\end{equation}
\subsubsection*{Representing Data as Paper and Funding Impact} \quad \label{sec:imp}
We explored an alternative representation of the scientific activity ( $g^P$ ) as a quartile-quantized timeseries of the \textit{total citation count of all papers} on the topic or its children topics in MeSH. We explored an alternative representation of  the industrial activity ( $g^S$ ) as a quartile-quantized timeseries of the total funding amount (in dollars) allocated to small businesses working on those topics, or their children topics in MeSH. In Equation 3  below, we formally denote how $g^P$ and $g^S$ are generated from the PubMed and SBIR data: 
\begin{equation} \tag{4} 
g^X(t, m) = \frac{1}{{K^X(t)}} \sum_{j \in X(t)} \mathcal{Q}_m \left[  c(j) \, H^j(m) \right] 
\end{equation}

\begin{figure}[t!]
  \centering
  \includegraphics[width=1\linewidth]{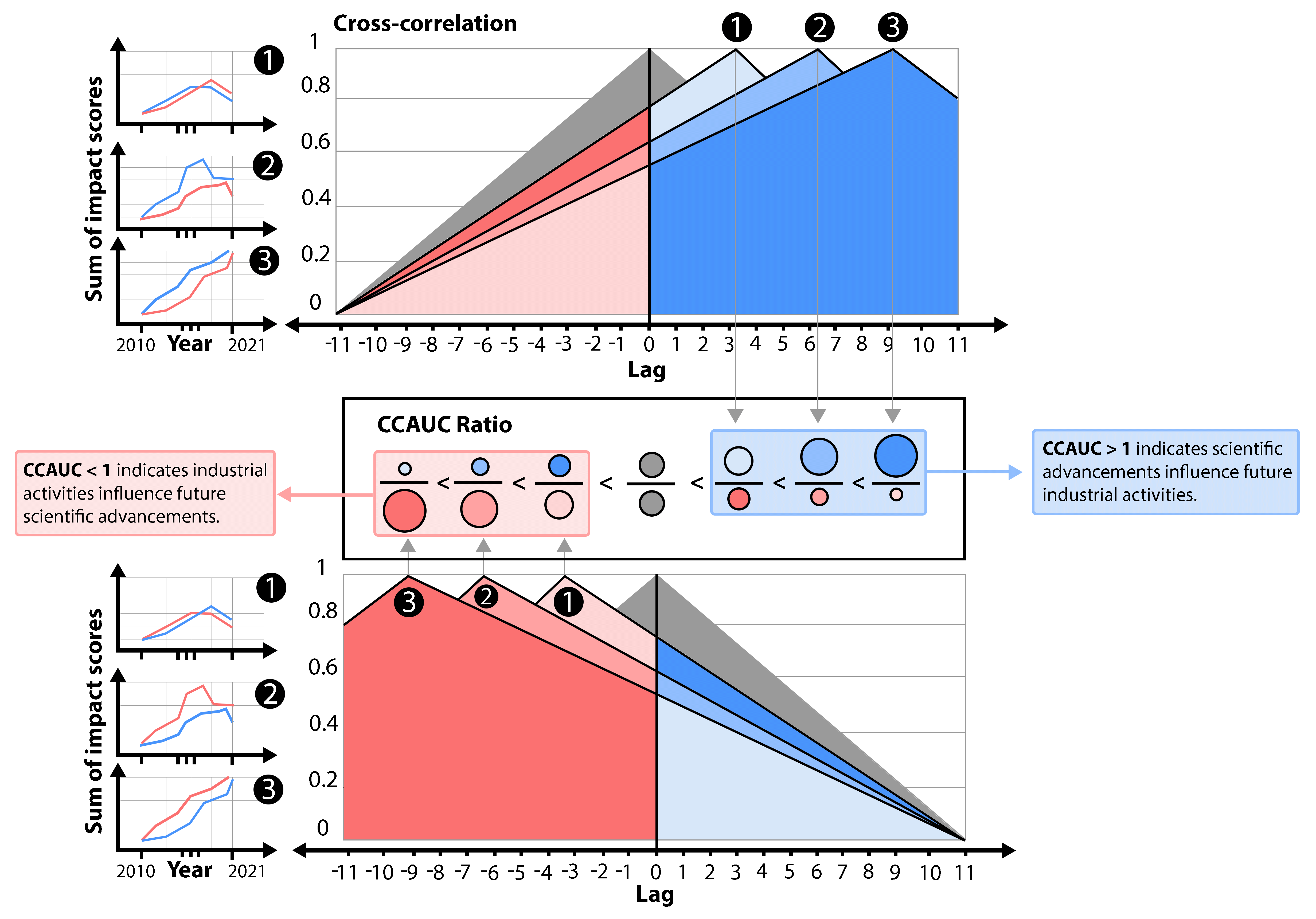}
  \caption{This figure illustrates the calculation of the \textit{Cross-Correlation Area Under the Curve} (CCAUC) ratio, a measure of the lead-lag relationship between scientific advancements and industrial activities. The line graphs on the left side, with science represented by blue lines and industry by red lines, track the sum of impact scores over time (note that the line graphs are schematic representations and do not depict actual trends; they are included solely for illustration purposes to demonstrate the conceptual framework). In the top cross-correlation plot, a rightward skew with a larger blue area indicates scenarios where scientific trends precede industrial ones, yielding a greater CCAUC ratio. The bottom plot shows the inverse, with a leftward skew and a more pronounced red area where industry leads science. The CCAUC ratio itself is derived by dividing the positive-lagged area (science leading) by the negative-lagged area (industry leading) under the cross-correlation curve. This distinction between line graphs and cross-correlation plots highlights not only the direction but also the temporal lag and the quantitative extent of impact between scientific research and industrial application.} 
  \label{fig:ccauc}
\end{figure}

\noindent Where $X \in \{P, S\}$; $c(j)$ is a function that returns the total number of citations if $j$ is a scientific abstract, and the award amount if it is a project description. Given $\mathcal{A}$ as $m$'s parent at the highest level of the hierarchy, $\mathcal{Q}_m$ is a function that converts the raw citation count to its corresponding quartile position when considering the distribution of citations within a given time-step, across topics $\{\mathcal{A}, \mathcal{A}_c\}$, Lastly,
\begin{equation} \tag{5}
{K^X(t)} = \sum_{j \in X(t)} \mathcal{Q}_m \left[ c(j) \right]  
\end{equation}
\noindent is a citation normalization constant where $m$ is the topic of abstract $j$. $g^{P}$ and $g^{S}$ were further normalized within each topic (i.e. across time) as $\tilde{g}^{P}$ and $\tilde{g}^{S}$ using Min-Max scaling following the same approach described in Equation 3 above.

\subsubsection*{Measurement of trend association}  
\label{sec:cc}
Cross-correlation (CC) is a measure of similarity between two signals as a function of a time-lag applied to one of them. In the context of this paper, cross-correlation was applied to each pair of scientific and industrial representations defined in Sections "\hyperref[sec:freq]{Representing Data as Paper and SBIR Award Frequency}" and "\hyperref[sec:imp]{Representing Data as Paper and Funding Impact}". More specifically, for all topics, we computed the cross correlations $\tilde{f}^P(t,m) * \tilde{f}^S(t-\tau,m)$, and $\tilde{g}^P(t,m) * \tilde{g}^S(t-\tau,m)$ where $\tau$ represents the number of years the industrial signal was shifted and varied from -11 to 11 (inclusive).

Lags in CC analysis are crucial for understanding temporal relationships between trends. For instance, if the industrial frequency trend ($\tilde{f}^{S}$) for a specific MeSH term peaks a few years after the same trend in the scientific domain ($\tilde{f}^{P}$) suggests a pattern where industrial trends are informed by prior scientific work in that term.

For each topic, we computed a single measure that denoted if the scientific activity was more likely to be leading industrial activity than vice versa. This measure was the ratio of: (1) the cumulative cross correlation for all positive $\tau$ and the cumulative cross correlation for all negative $\tau$. In Equations 6 and 7, we formally denote how this \textit{Cross-Correlation Area Under the Curve} (CCAUC) ratio was computed for the signal representation pairs defined in Sections "\hyperref[sec:freq]{Representing Data as Paper and SBIR Award Frequency}" and "\hyperref[sec:imp]{Representing Data as Paper and Funding Impact}" respectively:
\begin{equation} \tag{6}
\text{CCAUC}_f(m) = \frac
{1 + \sum_{\tau = 0}^{11} \tilde{f}^P(t,m) * \tilde{f}^S(t-\tau,m) }
{1 + \sum_{\tau = -11}^{0} \tilde{f}^P(t,m) * \tilde{f}^S(t-\tau,m)}
\end{equation}
\begin{equation} \tag{7}
\text{CCAUC}_g(m) = \frac
{1 + \sum_{\tau = 0}^{11} \tilde{g}^P(t,m) * \tilde{g}^S(t-\tau,m) }
{1 + \sum_{\tau = -11}^{0} \tilde{g}^P(t,m) * \tilde{g}^S(t-\tau,m)}
\end{equation} 

A CCAUC ratio exceeding 1 implies that the industrial trend was more likely to have lagged the scientific trend than vice versa. Conversely, a ratio below 1 suggests the reverse --- an industrial trend precluding its scientific counterpart. A ratio equal to 1 implies no time-lagged relationship between the trends. Thus, the CCAUC ratio provides us with a single measure to study if scientific activity was more likely to be leading industrial activity. In Figure \ref{fig:ccauc}, we provide an illustrative depiction of the CCAUC Ratio. In essence, the CCAUC ratio indicates the temporal delay and the directional correlation between scientific discoveries and industrial applications, highlighting the sequence and magnitude of their correlation. It offers a nuanced view of the chronological interconnection that shapes the trajectory of advancements across these domains.

\subsubsection*{Assessment of topic hierarchy on trend association} \label{sec:assess}
Our study acknowledges the complexity inherent in translating scientific discoveries into industrial applications, which varies not only in pace but also in the level of detail. To accurately reflect this diversity, we adopt a hierarchical analysis approach that examines the interplay between scientific and industrial trends across different levels of the MeSH taxonomy. By progressively navigating from broader categories to more specialized ones, we aim to illuminate the varying degrees of influence that scientific research exerts on industrial activity, from general trends to niche advancements. Given that the MeSH tree is composed of 13 layers, our traversal process involved a step-wise descent into each successive breadth level. Within each level, we incorporated all topics that are nested from the tree's root to our current depth. For every topic, denoted as $m$, we calculated CCAUC$(m)$ and determined the proportion of these values that exceeded 1, as well as those equal to or less than 1. Furthermore, we computed the Maximum Cross-Correlation (MCC) lag for each topic. This enabled us to determine the temporal change at which the correlation between scientific and industrial trends of $m$ reached its maximum prominence.

To establish confidence intervals at each depth, we iteratively computed the CCAUC ratios for various subsets of scientific and industrial trends. Using a sliding window parameter, $\mathfrak{w}$, which ranges from 1 to 11 years, we strategically select subsets of time series, $f^{X}_{2021-\mathfrak{w} : 2021}$ and $g^{X}_{2021-\mathfrak{w} : 2021}$. This selection process enables us to observe the evolution of the CCAUC ratio distribution in relation to varying values of $\mathfrak{w}$. Consequently, the error bounds computed are representative of the standard deviation of the corresponding CCAUC ratios at each depth level. 

\subsubsection*{Measurement of trend causality} \label{sec:granger}
Granger Causality (GC) is a statistical approach that assesses whether changes in one time series can predict changes in another. It's particularly useful in understanding potential relationships between two evolving trends across different lags. In this research, we applied GC to investigate the connections between scientific and industrial domains, as described in Sections "\hyperref[sec:freq]{Representing Data as Paper and SBIR Award Frequency}" and "\hyperref[sec:imp]{Representing Data as Paper and Funding Impact}". For each topic, we utilized the chi-squared GC test to determine (1) the extent to which current normalized annual frequencies of scientific papers ($\tilde{f}^P$) can be indicative of future normalized frequencies of SBIR grants ($\tilde{f}^S$) associated with the same topic, and (2) the degree to which the current normalized citation counts of scientific papers ($\tilde{g}^P$) serve as predictors for subsequent cumulative funding allocated through SBIR grants for that topic ($\tilde{g}^S$). The outcome of this test, represented as a p-value, elucidates the statistical significance of the causal relationships between the paired signals.

\subsubsection*{Assessment of topic hierarchy on trend causality} \label{sec:assess_granger}
Leveraging the hierarchical methodology from Section "\hyperref[sec:assess]{Assessment of topic hierarchy on trend association}", we traversed the MeSH tree to measure GC between scientific and industrial trends. For each topic, $m$, we determined the causality's significance by calculating its p-value and gauged the proportion below 0.05. Employing time lags up to 11 years for each analysis, we traced the distribution of GC significance across MeSH depths. The computed error bounds reflect the yearly standard deviation of these significant p-value ratios for each layer.

\subsection*{Methods for Research Question 2}\label{sec:contextanalysis}
For a given topic in the biomedical sciences, the goal of our second research question is to understand how the \textit{historical content} of scientific abstracts can be a leading indicator of the \textit{future content} of innovation grant applications for small businesses working on those topics. 

\subsubsection*{Representing the data as text embeddings } \label{sec:repemb} \quad
To answer our second research question, we investigated temporal associations between the semantic content of scientific papers and SBIR abstracts. More specifically, for all PubMed and SBIR abstracts in a given topic, we utilized the E5 (\urlstyle{same}\url{https://huggingface.co/intfloat/e5-large}) embedding model \cite{wang2022text} recognized for its state-of-the-art performance in text representation, to generate text embeddings, and studied their associations over time. Formally, given MeSH term $m$ and ($y_p$, $y_s$) as a pair of years ranging from 2010 to 2021, we select all scientific and industrial papers labeled with $m$ in years $y_p$ and $y_s$, respectively. We then embed the corresponding scientific abstracts and industrial project descriptions of those papers into a 1024-dimensional space. Next, we transform these high-dimensional embeddings into a two-dimensional space using Principal Component Analysis (PCA) and Uniform Manifold Approximation and Projection (UMAP) \cite{McInnes2018} --- noted for its efficiency with large datasets and preservation of global structure compared to t-SNE \cite{vandermaaten08a} --- for capturing linear and non-linear structures inherent in high-dimensional embeddings, respectively. We discretize this two-dimensional space, setting the number of bins along each dimension to $b_{1,2}=$ 20, yielding a 20x20 matrix. Each cell of this matrix represents a unique region in the semantic context space. 

\begin{figure}[t]
  \centering
  \includegraphics[width=0.7\linewidth]{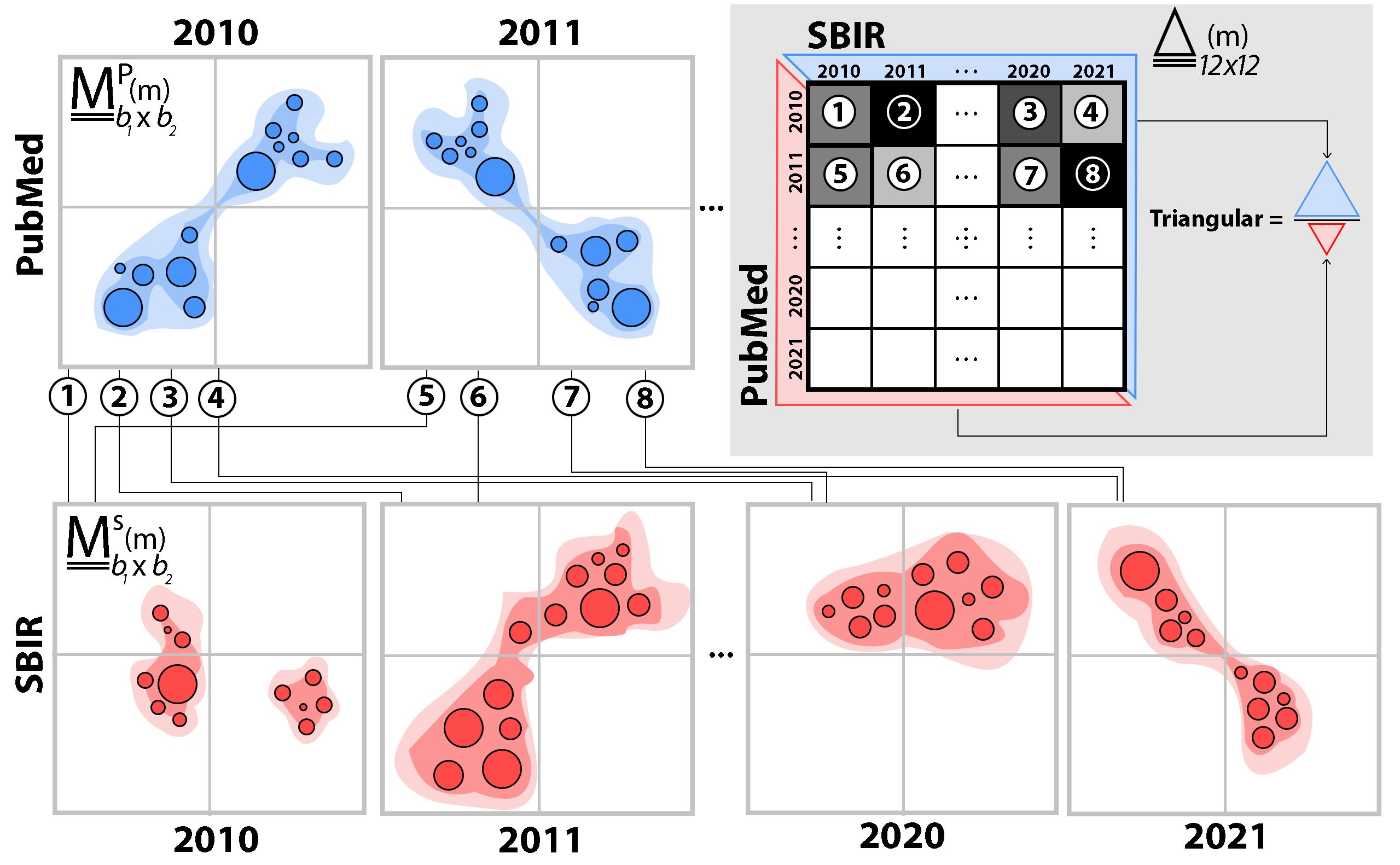}
  \caption{For each year pair, scientific and industrial abstracts are spatially mapped and normalized for similarity calculations. Each pair of embeddings generates a unique similarity value for its respective position in the grid. The contextual similarity between current scientific abstracts and future industrial project descriptions populates in the upper triangle, and vice versa in the lower triangle. The triangular ratio (\( tr \)), representing the influence of scientific context on future industrial projects, is the cumulative sum of similarities in the upper to lower triangle.}
  \label{fig:triangular}
\end{figure}

For each bin in the grid, we sum the total quartile-quantized citations of all scientific abstracts and the total quartile-quantized award amounts of all industrial abstracts that fall into it using the approach from Section "\hyperref[sec:imp]{Representing Data as Paper and Funding Impact}". Given $D_{xy}$ as the set of embedding points discretized at bins x and y of the $b_1 \times b_2$ grid, we calculate the density of that point for scientific and industrial abstracts using Equation 8:
\begin{equation} \tag{8}
\underline{\underline{M}}^{X}_{\mathrlap{b_1\times b_2}}(t, m) = \sum_{j\in d_{xy}} \mathcal{Q}_{m}[c(j)] 
\label{eq:msn}
\end{equation}
\noindent Where: $X \in \{P, S\}$; $x,y \in \mathbb{Z}$, $1 \leq x \leq b_1$, $1 \leq y \leq b_2$.

To smooth the densities, we apply Gaussian Kernel Density Estimation to the grid (we experimentally set the bandwidth to 0.8). The density values are then normalized to range from 0 to 1, creating a pseudo-probabilistic distribution for the semantic contexts present in the abstracts for each MeSH term within each domain. 

Next, we calculate the distance between the two probability distributions of the scientific and industrial contexts using the Total Variational Distance and the Hellinger distance. We then subtract the distances from 1 to measure the similarity between the two context distributions, effectively quantifying the degree of semantic overlap between the scientific and industrial abstracts for a given MeSH term. The aforementioned steps are repeated per topic for all possible pairs of ($y_1$,$y_2$), resulting in a 12x12 matrix of similarity scores, which we refer to as $\underline{\underline{\Delta}}_{\mathrlap{12 \times 12}}(m)$. 

\subsubsection*{Measurement of the content association} \label{sec:trian} \quad
For each topic, we computed a single measure that denoted if the content of scientific abstracts were more likely to be leading the content of industrial abstracts than vice versa. We denote this measurements as the \textit{triangular ratio} (\( tr \)) and formally define it in Equation \ref{eq:ss}.
\begin{equation} \tag{9}
\text{tr}(\underline{\underline{\Delta}}(m)) = \frac{1 + \sum_{\substack{i, j \in \{1,\dots,12\} \\ \mathbf{i<j}}} \underline{\underline{\Delta}}(m)[i, j]}{1 + \sum_{\substack{i, j \in \{1, \dots, 12\} \\ \mathbf{i\geq j}}} \underline{\underline{\Delta}}(m)[i, j]}
\label{eq:ss}
\end{equation}
Here, the numerator (upper triangular) compiles the sum of elements where scientific abstracts are posited to influence industrial counterparts at subsequent time points, while the denominator (lower triangular) aggregates the elements reflecting the opposite—industrial influence on science. The ratio \( tr \) thereby provides a concise metric of the directional disparity in the distribution of content themes: a value greater than 1 suggests a trend where scientific advancements inform industrial activities (\( tr > 1 \), indicating industry lags science), and conversely, a value less than 1 points to industrial activities informing scientific advancements (\( tr < 1 \), indicating science lags industry). The visualization of these content associations is detailed in Figure \ref{fig:triangular}. 

\begin{figure}[t]
  \centering
  \includegraphics[width=0.8\linewidth]{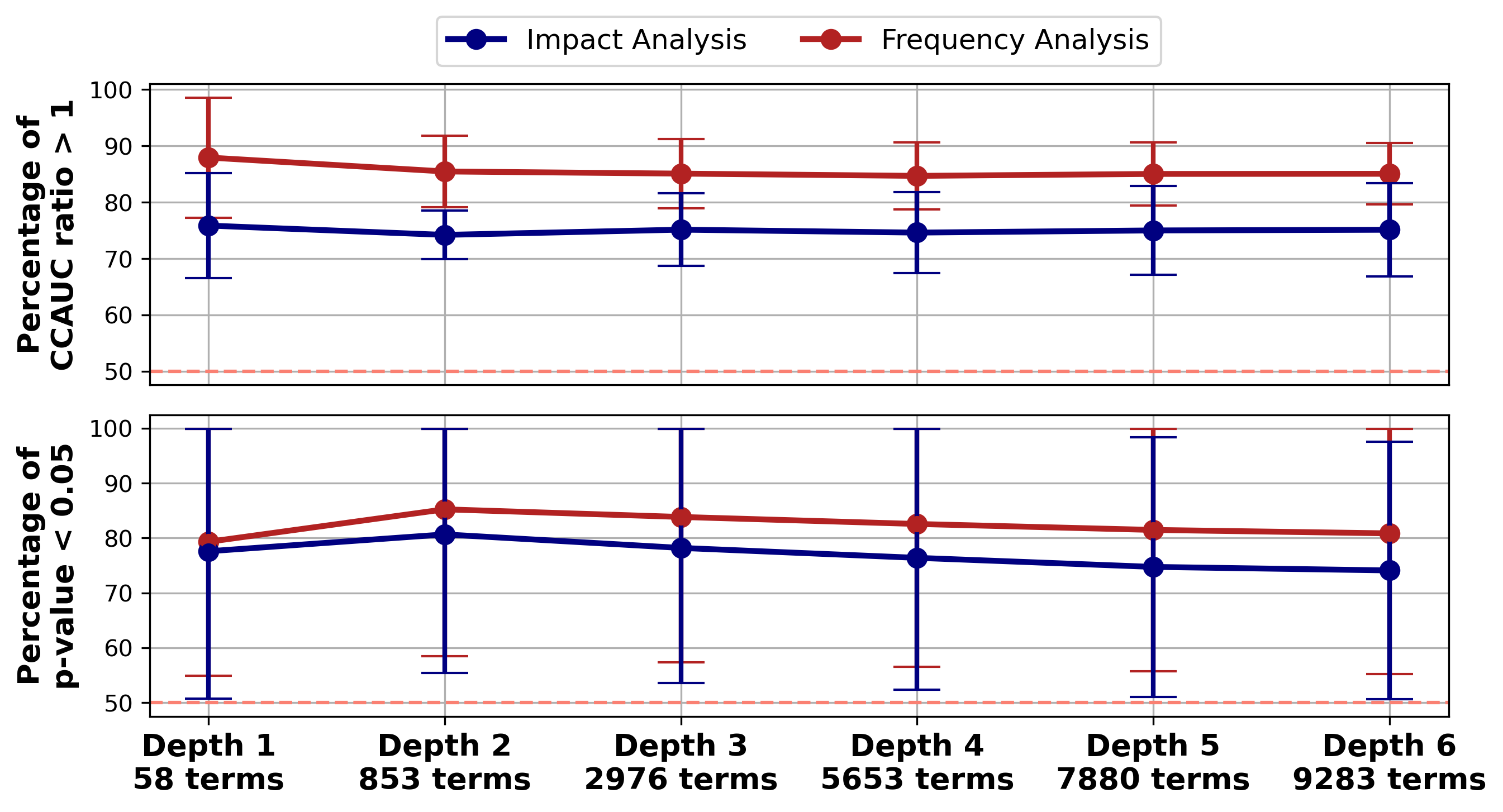}
  \caption{The precentage of CCAUC ratios exceeding one (see "\hyperref[sec:cc]{Measurement of trend association}") (top) and significant p-values (bottom) for both frequency (see "\hyperref[sec:freq]{Representing Data as Paper and SBIR Award Frequency}") and impact (see "\hyperref[sec:imp]{Representing Data as Paper and Funding Impact}") representations of the scientific and industrial data at multiple depths of MeSH terms. The x-axis denotes the topic resolution within the MeSH hierarchy. The error bar represents the deviation of this ratio as we shrink the window size ($\mathfrak{w}$).} 
  \label{fig:imp_vs_freq}
\end{figure}
\subsubsection*{Assessment of topic hierarchy on content association}
We adopted a similar strategy to the frequency and impact analysis for the context analysis, using the \( tr \) ratio, rather than the CCAUC, to assess semantic congruence as we traversed the depth of the MeSH tree. Furthermore, we leveraged the sliding window parameter $\mathfrak{w}$ (refer to Section "\hyperref[sec:assess]{Assessment of topic hierarchy on trend association}") to systematically select subsets of the similarity matrix $\underline{\underline{\Delta}}_{12 - \mathfrak{w} \times 12 - \mathfrak{w}}$. This selection methodology facilitated the tracking of the evolution of the \( tr \) ratio distribution according to different values of $\mathfrak{w}$. As such, the calculated error bounds reflect the standard deviation of the corresponding \( tr \) ratios at each depth level.

\begin{figure}[t]
  \includegraphics[width=1\linewidth]{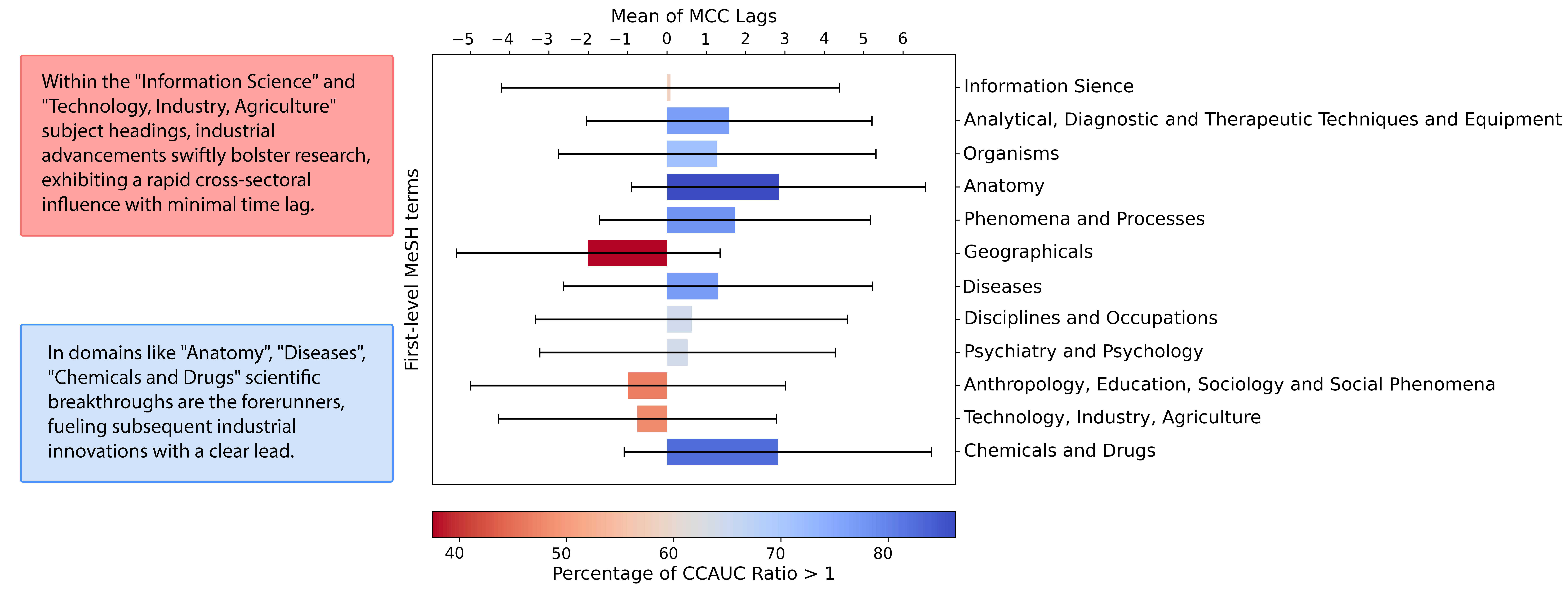}
  \caption{The Mean MCC lag (see "\hyperref[sec:assess]{Assessment of topic hierarchy on trend association}") for the impact representation of the scientific and industrial data, decomposed by the topics at the first level of the MeSH hierarchical tree. The MCC lag signifies the time delay at which science and industry trends are most correlated. The lengths of the bars correspond to the average MCC lag for each term and its child terms. MCC lags between 0 to 11 indicate that science impacts industry the most at that delay, while lags ranging from 0 to -11 indicate the vice versa influence. The colors of the bars represent the proportion of greater-than-one CCAUC ratios among a topic and its children topics. A higher proportion (cooler colors) implies a science-to-industry influence for the majority of the child terms, while a lower proportion (warmer colors) suggests an industry-to-science impact.}
  \label{fig:first_level_mcc_lags}
\end{figure}

\subsection*{Interdisciplinary Research and Innovation}
We also investigated the research questions in Section "\hyperref[sec:rques]{Research Questions}" among scientific studies and small businesses focusing on interdisciplinary topics. As scientific and industrial abstracts are labeled with sets of MeSH topics, we shifted our methodology from identifying these abstracts as instances of individual MeSH topics to labeling them based on pairs of these topics. Given the vast number of possible pairs, we applied the Pareto principle and selected the top 20\% most frequent pairs, amounting to approximately 35k pairs. Subsequently, we carried out frequency, impact, and context analyses as outlined in Sections "\hyperref[sec:freqimpactanalysis]{Methods for Research Question 1}" and "\hyperref[sec:contextanalysis]{Methods for Research Question 2}".

\section*{Results}
\subsection*{Results for Research Question 1}
In Figure \ref{fig:imp_vs_freq}, we present the results of trend analyses for both the frequency (see "\hyperref[sec:freq]{Representing Data as Paper and SBIR Award Frequency}") and impact (see "\hyperref[sec:imp]{Representing Data as Paper and Funding Impact}") representations of scientific and industrial data across various MeSH term depths, as illustrated in the top subfigure. The frequency trend analysis showed a stable percentage of CCAUC ratios greater than one, with approximately 88\% of topics at the primary depth (d=1). This percentage decreased slightly to 85\% as we delved deeper into the MeSH layers. While our analysis spanned all 13 MeSH levels, the figure visually represents up to the sixth depth for clarity, given that the percentage differences in levels 7 to 13 were minimal ($\pm$ 2\%). This analysis revealed a consistently strong influence of contemporary scientific activities on future industrial projects across different MeSH topic depths. This notable observation not only illustrates the inherent characteristic of frequency modeling but also validates the presumption that most scientific advancements will eventually find a commercial application among small businesses. This is especially pertinent since we selected the set of MeSH topics that had at least one industrial project granted based on that topic. This implies a broad and general interest from industry in scientific outcomes, which is reflected across different research depth layers. The results for the impact trend, on the other hand, indicated a percentage that ranged between 76\% to 74\% across various depths, demonstrating a slightly decreasing yet stable influence as we navigated deeper into the MeSH hierarchy. Similarly to the frequency trend, the impact trend remained consistent across different depth levels, suggesting a steady influence of science on industry, which holds even with the increasing volume of scientific outputs. Crucially, this influence indicates that industrial \textit{funding} is not merely driven by the \textit{quantity} of scientific production. Instead, it underscores the industry's appreciation for the applicability and potential innovation stemming from \textit{impactful} scientific findings.

The GC test, depicted in the bottom subfigure, further elucidated the influence of scientific advancements on industrial activities. For the frequency trend analysis, about 80\% of MeSH topics across different depths consistently exhibited p-values less than 0.05, underscoring the significant predictive power of contemporary scientific activities on future industrial projects. In contrast, the impact trend analysis displayed a moderate range from 78\% to 73\% for significant p-values across the MeSH hierarchy. This suggests that while there's a predominant influence of impactful scientific activities on the industrial domain, the intensity of this influence experiences a slight tapering as we traverse deeper into the MeSH layers.

In Figure \ref{fig:first_level_mcc_lags}, we reveal the direction and time lag between scientific and industrial trends for the top level of the MeSH hierarchy. The analysis implies that the latency of science's impact on industrial funding varies significantly by topic. In particular, domains like "Anatomy" and "Chemicals and Drugs" exhibit a notably extended latency in the influence of scientific discoveries on industry, as compared to fields such as 'Information Science,' where the data suggests a reciprocal influence with industrial trends potentially shaping scientific research. These findings highlight the complex and nuanced interplay between scientific inquiry and industrial application across different biomedical disciplines.

\begin{figure}[t]
  \centering
  \includegraphics[width=0.8\linewidth]{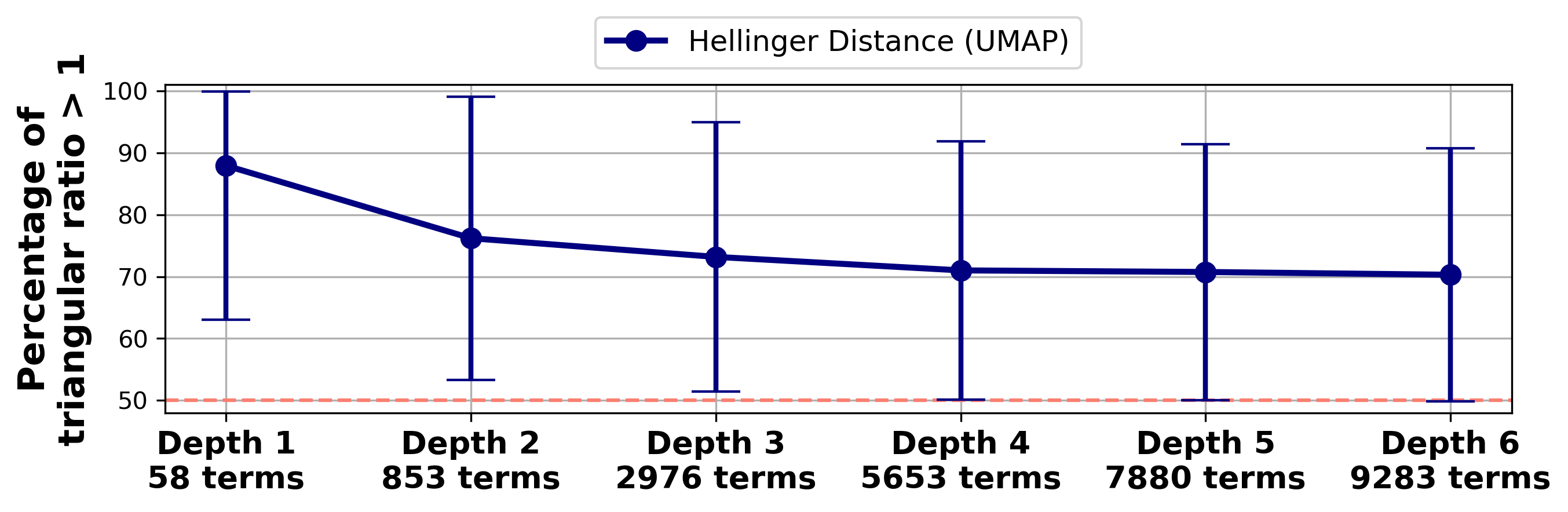}
  \caption{The precentage of triangular (\( tr \)) ratios exceeding one (see "\hyperref[sec:trian]{Measurement of the content association}") for contextual text embedding (see "\hyperref[sec:freq]{Representing Data as Paper and SBIR Award Frequency}") representations of the scientific and industrial data at multiple depths of MeSH terms. The x-axis denotes the topic resolution within the MeSH hierarchy, while the y-axis indicates the proportion of \( tr \) ratios in that topic set exceeding one. The error bar represents the deviation of this ratio as we shrink the window size ($\mathfrak{w}$).} 
  \label{fig:tvd_jsd_hld}
\end{figure}

\subsection*{Results for Research Question 2} 
In Figure \ref{fig:tvd_jsd_hld}, we present our analysis on the temporal associations between the semantic content of scientific papers and SBIR abstracts. We utilized UMAP and the Hellinger distance to reduce the dimensionality of text representations and to assess the similarity between pairs of semantics-based probability distributions. Our results underscore a notable degree of science-to-industry influence, evidenced by approximately 88\% of the \( tr \) ratios exceeding one at the first depth and exhibits a downward trend as we delve deeper into the MeSH tree, dropping to 70\%. Although not depicted in the figure, it important to mention that both the Total Variational Distance (TVD) and Hellinger distance (HD) for PCA, in addition to the TVD for UMAP, begin from the same 88\% benchmark. As depth increases, these metrics settle within a range from 70\% to 75\%, with a variation of about +5\%. Depths 7 to 13 showed results consistent with the fifth depth, deviating by only ($\pm$ 1\%) across all measures. This downward pattern potentially stems from the increasing specificity of the scientific abstracts deeper into the tree, resulting in less semantic overlap with the generally broader industrial abstracts. This analysis underscores the persistent and nuanced \textit{semantic impact} of \textit{current, impactful} scientific advancements on \textit{future} industrial funding. 

\subsection*{Results for Interdisciplinary Studies}
The incorporation of interdisciplinary studies into our analysis yielded compelling findings. Using frequency, impact, and contextual analyses, we identified a substantial science-to-industry influence. This was evidenced by CCAUC ratios greater than one in 82.5\% of topics for frequency analysis and 76\% for impact analysis, along with a 70\% greater-than-one \( tr \) ratio for the contextual analysis. Although depth-wise investigation is not applicable due to the diverse origins of high-frequency terms, the robust correlations underscore the substantial influence of interdisciplinary scientific advancements in shaping industry.

\section*{Discussion}
\subsection*{Key Findings}
\subsubsection*{Science as a leading indicator for industrial innovation funding:} Our analyses reveal that topics with the most influential scientific activities (i.e. those with more citations) are also the most likely to see future allocations of industrial funds. More specifically, for up-to 76\% of topics investigated, the scientific interest in papers from those topics were associated with the total funding allocated to small businesses working on those topics in the future. This result provides evidence that science informs industrial innovation funding decisions.

\subsubsection*{Science as a leading indicator for industrial innovation topics:} Our analysis reveals that the semantic contents of scientific abstracts within a topic are associated with the future semantic contents of grant applications of small businesses working on those topics. More specifically, for approximately 75.6\% of topics examined, text embeddings of scientific abstracts were associated with future industrial text embeddings. These findings prove that science influences the direction of industrial innovation activities within topics. 

\subsection*{Impact of Science on Industry}
The primary objective of our analysis was to investigate how current scientific advancements impact future industrial innovation. The frequency, impact, and context analyses provided multifaceted insights into the dynamics of this process, revealing how scientific progress influences the allocation of industrial funding across different thematic depths and contexts. The frequency analysis demonstrated that various scientific activities significantly inform future industrial projects across varying depths of MeSH topic categorization. This pattern indicates a broad, general interest from industry in scientific outcomes, with a substantial portion of scientific discourse finding its way into commercial application. Meanwhile, the impact analysis refined this understanding by modeling the evolution of science and industry by capturing the level of interest each work attracted in its respective field, showing substantial influence in guiding industrial investments and directions, regardless of the sheer volume of scientific production. This balance between quantity and quality elucidates a nuanced interplay between scientific research and industrial innovation, reaffirming the pivotal role that impactful science plays in shaping industrial progress. Shifting the focus to context analysis, it accentuated the profound semantic influence of scientific discourse on industrial innovation. However, the deeper we went into the MeSH tree, the more specific and technical language we encountered in scientific abstracts. However, SBIR abstracts tend to maintain a more general vocabulary, which often doesn't reflect these highly specific terms. This semantic divergence leads to a decrease in the percentage of terms with a \( tr \) ratio above one, indicating a lesser degree of semantic overlap as we move to more specific terms. Consequently, our study portrays the significant influence of science on the industry, emphasizing that this relationship extends beyond the mere volume of scientific output, but is greatly influenced by its impact, the broader themes it advances, and the meaningful narratives it presents, which collectively underline the central role of science in steering industrial innovation.

\subsection*{Limitations}
Our study offers insights into the science-industry relationship but has some limitations. The ontology used, while extensive, only represents a specific scientific domain; a broader ontology could yield deeper insights. Expanding the study's timespan might refine error bounds through advanced statistical methods. While our data source is robust, larger databases might offer further insights. This study centers on analyzing science-industry associations, but future work could predict upcoming trends in this interplay. Nonetheless, our findings set a solid groundwork for subsequent research.

\section*{Conclusion}
This study aimed to model the interaction between scientific research and industrial innovation using different techniques, including frequency, impact, and context analysis. Our results consistently underscored scientific advancements' decisive role in shaping industrial innovation. Influential scientific activities substantially align with future industrial funding, and the thematic content of scientific discourse profoundly affects industrial innovation. These findings illuminate the nuanced interplay between science and industry, which is dictated by not only the quantity of scientific output but also its relevance and impact. Future research can capitalize on our findings while also addressing the limitations outlined herein. Such work would contribute to a greater understanding of the intricate dynamics between scientific exploration and industrial innovation.

\section*{Acknowledgements}
This research was funded in part by the Faculty Research Awards of J.P. Morgan AI Research. The authors are solely responsible for the contents of the paper and the opinions expressed in this publication do not reflect those of the funding agencies. 

\textbf{Disclaimer} This paper was prepared for informational purposes by the Artificial Intelligence Research group of JPMorgan Chase \& Co and its affiliates (“JP Morgan”), and is not a product of the Research Department of JP Morgan. JP Morgan makes no representation and warranty whatsoever and disclaims all liability, for the completeness, accuracy or reliability of the information contained herein. This document is not intended as investment research or investment advice, or a recommendation, offer or solicitation for the purchase or sale of any security, financial instrument, financial product or service, or to be used in any way for evaluating the merits of participating in any transaction, and shall not constitute a solicitation under any jurisdiction or to any person, if such solicitation under such jurisdiction or to such person would be unlawful.

\bibliography{sample}

\begin{thebibliography}{10}
\urlstyle{rm}
\expandafter\ifx\csname url\endcsname\relax
  \def\url#1{\texttt{#1}}\fi
\expandafter\ifx\csname urlprefix\endcsname\relax\def\urlprefix{URL }\fi
\expandafter\ifx\csname doiprefix\endcsname\relax\def\doiprefix{DOI: }\fi
\providecommand{\bibinfo}[2]{#2}
\providecommand{\eprint}[2][]{\url{#2}}

\bibitem{Gopukumar}
\bibinfo{author}{Luo, J.}, \bibinfo{author}{Wu, M.}, \bibinfo{author}{Gopukumar, D.} \& \bibinfo{author}{Zhao, Y.}
\newblock \bibinfo{journal}{\bibinfo{title}{Big data application in biomedical research and health care: A literature review}}.
\newblock {\emph{\JournalTitle{Biomedical Informatics Insights}}} \textbf{\bibinfo{volume}{8}}, \bibinfo{pages}{1}, \doiprefix\url{10.4137/BII.S31559} (\bibinfo{year}{2016}).

\bibitem{Jinek}
\bibinfo{author}{Jinek, M.} \emph{et~al.}
\newblock \bibinfo{journal}{\bibinfo{title}{A programmable dual-rna-guided dna endonuclease in adaptive bacterial immunity}}.
\newblock {\emph{\JournalTitle{Science (New York, N.Y.)}}} \textbf{\bibinfo{volume}{337}}, \bibinfo{pages}{816--21}, \doiprefix\url{10.1126/science.1225829} (\bibinfo{year}{2012}).

\bibitem{Mounadi}
\bibinfo{author}{El~Mounadi, K.}, \bibinfo{author}{Morales-Floriano, M.} \& \bibinfo{author}{Garcia-Ruiz, H.}
\newblock \bibinfo{journal}{\bibinfo{title}{Principles, applications, and biosafety of plant genome editing using crispr-cas9}}.
\newblock {\emph{\JournalTitle{Frontiers in Plant Science}}} \textbf{\bibinfo{volume}{11}}, \doiprefix\url{10.3389/fpls.2020.00056} (\bibinfo{year}{2020}).

\bibitem{shammas2011telomeres}
\bibinfo{author}{Shammas, M.~A.}
\newblock \bibinfo{journal}{\bibinfo{title}{Telomeres, lifestyle, cancer, and aging}}.
\newblock {\emph{\JournalTitle{Current opinion in clinical nutrition and metabolic care}}} \textbf{\bibinfo{volume}{14}}, \bibinfo{pages}{28--34}, \doiprefix\url{10.1097/MCO.0b013e32834121b1} (\bibinfo{year}{2011}).

\bibitem{N2023}
\bibinfo{author}{for Science, N.~C.} \& \bibinfo{author}{Statistics, E.}
\newblock \bibinfo{title}{Federal budget authority for r\&d and r\&d plant for national defense and civilian functions totaled \$191 billion in fy 2023 proposed budget}.
\newblock \bibinfo{howpublished}{https://ncses.nsf.gov/pubs/nsf23323} (\bibinfo{year}{2023}).
\newblock \bibinfo{note}{Accessed on January 26, 2023}.

\bibitem{Jürgens}
\bibinfo{author}{Jürgens, B.} \& \bibinfo{author}{Herrero-Solana, V.}
\newblock \bibinfo{journal}{\bibinfo{title}{Patent bibliometrics and its use for technology watch}}.
\newblock {\emph{\JournalTitle{Journal of Intelligence Studies in Business}}} \textbf{\bibinfo{volume}{7}}, \bibinfo{pages}{17--26}, \doiprefix\url{10.37380/jisib.v7i2.236} (\bibinfo{year}{2017}).

\bibitem{Skute}
\bibinfo{author}{Skute, I.}, \bibinfo{author}{Zalewska-Kurek, K.}, \bibinfo{author}{Hatak, I.} \& \bibinfo{author}{Weerd‐Nederhof, P.}
\newblock \bibinfo{journal}{\bibinfo{title}{Mapping the field: A bibliometric analysis of the literature on university–industry collaborations}}.
\newblock {\emph{\JournalTitle{The Journal of Technology Transfer}}} \textbf{\bibinfo{volume}{44}}, \bibinfo{pages}{916--947}, \doiprefix\url{10.1007/s10961-017-9637-1} (\bibinfo{year}{2019}).

\bibitem{García}
\bibinfo{author}{Magadán~Díaz, M.} \& \bibinfo{author}{García, J.}
\newblock \bibinfo{journal}{\bibinfo{title}{Publishing industry: A bibliometric analysis of the scientific production indexed in scopus}}.
\newblock {\emph{\JournalTitle{Publishing Research Quarterly}}} \textbf{\bibinfo{volume}{38}}, \doiprefix\url{10.1007/s12109-022-09911-3} (\bibinfo{year}{2022}).

\bibitem{COBO2018364}
\bibinfo{author}{Cobo, M.}, \bibinfo{author}{Jürgens, B.}, \bibinfo{author}{Herrero-Solana, V.}, \bibinfo{author}{Martínez, M.} \& \bibinfo{author}{Herrera-Viedma, E.}
\newblock \bibinfo{journal}{\bibinfo{title}{Industry 4.0: a perspective based on bibliometric analysis}}.
\newblock {\emph{\JournalTitle{Procedia Computer Science}}} \textbf{\bibinfo{volume}{139}}, \bibinfo{pages}{364--371}, \doiprefix\url{https://doi.org/10.1016/j.procs.2018.10.278} (\bibinfo{year}{2018}).
\newblock \bibinfo{note}{6th International Conference on Information Technology and Quantitative Management}.

\bibitem{KRESTEL2021102035}
\bibinfo{author}{Krestel, R.}, \bibinfo{author}{Chikkamath, R.}, \bibinfo{author}{Hewel, C.} \& \bibinfo{author}{Risch, J.}
\newblock \bibinfo{journal}{\bibinfo{title}{A survey on deep learning for patent analysis}}.
\newblock {\emph{\JournalTitle{World Patent Information}}} \textbf{\bibinfo{volume}{65}}, \bibinfo{pages}{102035}, \doiprefix\url{https://doi.org/10.1016/j.wpi.2021.102035} (\bibinfo{year}{2021}).

\bibitem{su15043484}
\bibinfo{author}{Zhu, Y.}, \bibinfo{author}{Wang, Y.}, \bibinfo{author}{Zhou, B.}, \bibinfo{author}{Hu, X.} \& \bibinfo{author}{Xie, Y.}
\newblock \bibinfo{journal}{\bibinfo{title}{A patent bibliometric analysis of carbon capture, utilization, and storage (ccus) technology}}.
\newblock {\emph{\JournalTitle{Sustainability}}} \textbf{\bibinfo{volume}{15}}, \doiprefix\url{10.3390/su15043484} (\bibinfo{year}{2023}).

\bibitem{PULIGA2023}
\bibinfo{author}{Puliga, G.}, \bibinfo{author}{Bono, F.}, \bibinfo{author}{Gutierrez~Tenreiro, E.~G.} \& \bibinfo{author}{Strozzi, F.}
\newblock \bibinfo{title}{Bibliometric analysis of scientific publications and patents on smart cities}.
\newblock \bibinfo{type}{Tech. Rep.} \bibinfo{number}{JRC129102}, \bibinfo{institution}{Publications Office of the European Union} (\bibinfo{year}{2023}).
\newblock \doiprefix\url{10.2760/074691}.

\bibitem{Zexia}
\bibinfo{author}{Wang, L.} \& \bibinfo{author}{Li, Z.}
\newblock \bibinfo{journal}{\bibinfo{title}{Knowledge flows from public science to industrial technologies}}.
\newblock {\emph{\JournalTitle{The Journal of Technology Transfer}}} \bibinfo{pages}{1--24}, \doiprefix\url{10.1007/s10961-019-09738-9} (\bibinfo{year}{2021}).

\bibitem{Chakraborty2020PatentCN}
\bibinfo{author}{Chakraborty, M.}, \bibinfo{author}{Byshkin, M.} \& \bibinfo{author}{Crestani, F.~A.}
\newblock \bibinfo{journal}{\bibinfo{title}{Patent citation network analysis: A perspective from descriptive statistics and ergms}}.
\newblock {\emph{\JournalTitle{PLoS ONE}}} \textbf{\bibinfo{volume}{15}}, \doiprefix\url{https://doi.org/10.1371/journal.pone.0241797} (\bibinfo{year}{2020}).

\bibitem{Tan2022}
\bibinfo{author}{Tan, W.}, \bibinfo{author}{Jing, L.}, \bibinfo{author}{Wang, Y.} \& \bibinfo{author}{Li, W.}
\newblock \bibinfo{journal}{\bibinfo{title}{A global bibliometric analysis on kawasaki disease research over the last 5 years (2017-2021)}}.
\newblock {\emph{\JournalTitle{Frontiers in Public Health}}} \textbf{\bibinfo{volume}{10}}, \bibinfo{pages}{1075659}, \doiprefix\url{10.3389/fpubh.2022.1075659} (\bibinfo{year}{2022}).

\bibitem{Khalid}
\bibinfo{author}{Farooq, K.}, \bibinfo{author}{Ur~Rehman, S.}, \bibinfo{author}{Ashiq, M.}, \bibinfo{author}{Siddique, N.} \& \bibinfo{author}{Ahmad~Phd, S.}
\newblock \bibinfo{journal}{\bibinfo{title}{Bibliometric analysis of coronavirus disease (covid-19) literature published in web of science 2019-2020}}.
\newblock {\emph{\JournalTitle{Journal of Family and Community Medicine}}} \textbf{\bibinfo{volume}{28}}, \bibinfo{pages}{1--7}, \doiprefix\url{10.4103/jfcm.JFCM_332_20} (\bibinfo{year}{2021}).

\bibitem{Mardikoraem}
\bibinfo{author}{Mardikoraem, M.} \& \bibinfo{author}{Woldring, D.}
\newblock \bibinfo{journal}{\bibinfo{title}{Protein fitness prediction is impacted by the interplay of language models, ensemble learning, and sampling methods}}.
\newblock {\emph{\JournalTitle{Pharmaceutics}}} \textbf{\bibinfo{volume}{15}}, \doiprefix\url{10.3390/pharmaceutics15051337} (\bibinfo{year}{2023}).

\bibitem{Maghrabi}
\bibinfo{author}{Maghrabi, Y.}, \bibinfo{author}{Ashgar, M.}, \bibinfo{author}{Aljohani, S.}, \bibinfo{author}{Alqarni, R.} \& \bibinfo{author}{Baeesa, S.}
\newblock \bibinfo{journal}{\bibinfo{title}{Three decades of spine surgery research evolution in saudi arabia: A bibliometric analysis}}.
\newblock {\emph{\JournalTitle{Journal of Spine Practice (JSP)}}} \textbf{\bibinfo{volume}{2}}, \bibinfo{pages}{51--60}, \doiprefix\url{10.18502/jsp.v2i2.12627} (\bibinfo{year}{2023}).

\bibitem{Chandrasekar}
\bibinfo{author}{Rubini, S.}, \bibinfo{author}{Chandrasekar, K.}, \bibinfo{author}{Janen, T.} \& \bibinfo{author}{Sriskandarajah, N.}
\newblock \bibinfo{journal}{\bibinfo{title}{Water quality in northern province of sri lanka: A bibliometric analysis of publications 1960–2021}}.
\newblock {\emph{\JournalTitle{World Water Policy}}} \textbf{\bibinfo{volume}{n/a}}, \doiprefix\url{https://doi.org/10.1002/wwp2.12117} (\bibinfo{year}{2023}).

\bibitem{Mohana}
\bibinfo{author}{Mohana~Murali, S.}, \bibinfo{author}{Senthamarai~Kannan, K.} \& \bibinfo{author}{Samuel, M.}
\newblock \bibinfo{journal}{\bibinfo{title}{Bibliometric analysis of the scientific literature on human papillomavirus vaccine clinical trials: Analysis of pubmed database}}.
\newblock {\emph{\JournalTitle{National Journal of Community Medicine}}} \textbf{\bibinfo{volume}{14}}, \bibinfo{pages}{424--32}, \doiprefix\url{10.55489/njcm.140720232951} (\bibinfo{year}{2023}).

\bibitem{Audretsch2019KnowledgeBK}
\bibinfo{author}{Audretsch, D.~B.}, \bibinfo{author}{Link, A.} \& \bibinfo{author}{van Hasselt, M.}
\newblock \bibinfo{journal}{\bibinfo{title}{Knowledge begets knowledge: university knowledge spillovers and the output of scientific papers from u.s. small business innovation research (sbir) projects}}.
\newblock {\emph{\JournalTitle{Scientometrics}}} \textbf{\bibinfo{volume}{121}}, \bibinfo{pages}{1367 -- 1383} (\bibinfo{year}{2019}).

\bibitem{Hayter}
\bibinfo{author}{Hayter, C.} \& \bibinfo{author}{Link, A.}
\newblock \bibinfo{journal}{\bibinfo{title}{From discovery to commercialization: accretive intellectual property strategies among small, knowledge-based firms}}.
\newblock {\emph{\JournalTitle{Small Business Economics}}} \doiprefix\url{10.1007/s11187-021-00446-z} (\bibinfo{year}{2021}).

\bibitem{pubmed}
\bibinfo{title}{Pubmed}.
\newblock \bibinfo{howpublished}{Internet} (\bibinfo{year}{2022}).
\newblock \bibinfo{note}{[Accessed: 2022-12-01]}.

\bibitem{sbir}
\bibinfo{title}{{SBIR} awards data} (\bibinfo{year}{2022}).
\newblock \bibinfo{note}{[Accessed: 2022-12-01]}.

\bibitem{Neumann_2019}
\bibinfo{author}{Neumann, M.}, \bibinfo{author}{King, D.}, \bibinfo{author}{Beltagy, I.} \& \bibinfo{author}{Ammar, W.}
\newblock \bibinfo{title}{{ScispaCy}: Fast and robust models for biomedical natural language processing}.
\newblock In \emph{\bibinfo{booktitle}{Proceedings of the 18th {BioNLP} Workshop and Shared Task}}, \doiprefix\url{10.18653/v1/w19-5034} (\bibinfo{publisher}{Association for Computational Linguistics}, \bibinfo{year}{2019}).

\bibitem{Bodenreider}
\bibinfo{author}{Bodenreider, O.}
\newblock \bibinfo{journal}{\bibinfo{title}{The unified medical language system (umls): Integrating biomedical terminology}}.
\newblock {\emph{\JournalTitle{Nucleic acids research}}} \textbf{\bibinfo{volume}{32}}, \bibinfo{pages}{D267--70}, \doiprefix\url{10.1093/nar/gkh061} (\bibinfo{year}{2004}).

\bibitem{ABBASI2021104121}
\bibinfo{author}{Abbasi, A.}, \bibinfo{author}{Miahi, E.} \& \bibinfo{author}{Mirroshandel, S.~A.}
\newblock \bibinfo{journal}{\bibinfo{title}{Effect of deep transfer and multi-task learning on sperm abnormality detection}}.
\newblock {\emph{\JournalTitle{Computers in Biology and Medicine}}} \textbf{\bibinfo{volume}{128}}, \bibinfo{pages}{104121}, \doiprefix\url{https://doi.org/10.1016/j.compbiomed.2020.104121} (\bibinfo{year}{2021}).

\bibitem{wang2022text}
\bibinfo{author}{Wang, L.} \emph{et~al.}
\newblock \bibinfo{title}{Text embeddings by weakly-supervised contrastive pre-training} (\bibinfo{year}{2022}).
\newblock \eprint{2212.03533}.

\bibitem{McInnes2018}
\bibinfo{author}{McInnes, L.}, \bibinfo{author}{Healy, J.}, \bibinfo{author}{Saul, N.} \& \bibinfo{author}{Großberger, L.}
\newblock \bibinfo{journal}{\bibinfo{title}{Umap: Uniform manifold approximation and projection}}.
\newblock {\emph{\JournalTitle{Journal of Open Source Software}}} \textbf{\bibinfo{volume}{3}}, \bibinfo{pages}{861}, \doiprefix\url{10.21105/joss.00861} (\bibinfo{year}{2018}).

\bibitem{vandermaaten08a}
\bibinfo{author}{van~der Maaten, L.} \& \bibinfo{author}{Hinton, G.}
\newblock \bibinfo{journal}{\bibinfo{title}{Visualizing data using t-sne}}.
\newblock {\emph{\JournalTitle{Journal of Machine Learning Research}}} \textbf{\bibinfo{volume}{9}}, \bibinfo{pages}{2579--2605} (\bibinfo{year}{2008}).

\end{thebibliography}

\section*{Author contributions}
Conceptualization, methodology, investigation, modeling, validation, and manuscript writing were performed by R.K. and M.M.G.—Review and scientific editing were performed by S.K., C.H.S., T.A., I.B., A.N., M.M.G., and R.K.—Data processing was executed by R.K., M.M.G and T.K.—Project administration: M.M.G. All authors reviewed the manuscript.

\section*{Competing Interests}
The authors declare no competing interests.

\section*{Data Availability}
All data utilized in this research, including datasets from PubMed, SBIR, and MeSH, are publicly available and accessible. The implementation, including code and relevant files, can be found at the project's GitHub repository (\urlstyle{same}\url{https://github.com/HAAIL/science-impacts-industry}).
\end{document}